\documentstyle[aps,prl,multicol,epsfig]{revtex}
\begin{document}
\title{ Broken symmetry, excitons, gapless modes and topological excitations in Trilayer Quantum Hall systems }
\author{  Jinwu Ye  }
\address{ Department of Physics, The Pennsylvania State University, University Park, PA, 16802 }
\date{\today}
\maketitle
\begin{abstract}
 We study the interlayer coherent incompressible phase in Trilayer Quantum Hall
 systems (TLQH) at total filling factor $ \nu_{T}=1 $ from three approaches:
 Mutual Composite Fermion (MCF), Composite Boson (CB) and wavefunction approach.
 Just like in Bilayer Quantum Hall system, CB approach is  superior than
 MCF approach in studying TLQH with broken symmetry. The Hall and Hall drag resistivities
 are found to be quantized at $ h/e^{2} $. Two neutral gapless modes 
 with linear dispersion relations are identified and the ratio of the two velocities is close to $ \sqrt{3} $.
 The novel excitation spectra are classified into two classes: Charge neutral bosonic
 2-body bound states and Charge $ \pm 1 $ fermionic 3-body bound states.
 In general, there are two 2-body Kosterlize-Thouless (KT) transition temperatures
 and one 3-body KT transition. The Charge $ \pm 1 $ 3-body fermionic
 bound states may be the main dissipation source of transport measurements.
 The broken symmetry in terms of $ SU(3) $ algebra is studied. The structure
 of excitons and their flowing patterns are given. The coupling between the two Goldstone modes
 may lead to the broadening in the zero-bias peak in the interlayer correlated tunnelings
 of the TLQH.  Several interesting features unique to TLQH are outlined.
 Limitations of the CB approach are also pointed out.
\end{abstract}
\begin{multicols}{2}

\section{ Introduction}
  Quantum Hall Effect (FQHE) in multicomponent systems has attracted
  a lot of attention since the seminal work by Halperin \cite{bert}. These components
  could be the spins of electrons when the Zeeman coupling is very small or layer indices in
  multi-layered system.
  In particular, spin-polarized Bilayer Quantum Hall (BLQH) systems at total filling factor
  $ \nu_{T} =1 $ have been under enormous experimental \cite{supp,gold,hall,em} 
  and theoretical \cite{fer,wen,jap,yang,moon,rev,mcf,cbprl} investigations over the last decade.
  When the interlayer separation $ d $ is sufficiently large,
  the bilayer system decouples into two separate compressible $ \nu=1/2 $ layers.
  However, when $ d $ is sufficiently small, in the absence
  of interlayer tunneling, the system undergoes a quantum phase transition into
  a novel spontaneous interlayer coherent incompressible phase \cite{rev}.

   By treating the two layer indices as two pseudo-spin indices, 
   Girvin, Macdonald and collaborators \cite{yang,moon,rev} mapped the bilayer system
   into a Easy Plane Quantum Ferromagnet (EPQFM). They achieved the mapping
   by projecting the Hamiltonian of 
   the BLQH onto the Lowest Landau Level (LLL) and then using
   subsequent Hartree-Fock (HF) approximation and gradient expansion ( called LLL+HF in the following ).
   They explored many rich and interesting physical phenomena in this system.
   The low energy excitations above the ground state is given by an effective $ 2+1 $ dimensional $ XY $ model.
   In addition to the Goldstone mode, there are also 4 flavors of topological
   defects called " merons " which carry fractional 
   charges $ \pm 1/2 $ and also have $ \pm $ vorticity. They have logarithmic divergent 
   self energies and are bound into pairs
   at low temperature. The lowest energy excitations carry charge $ \pm e $ which
   are a meron pair with opposite vorticity and
   the same charge. There is a finite temperature Kosterlize-Thouless
   (KT) phase transition at $ T_{KT} $ where
   bound states of the 4 flavors of merons are broken into free merons. Unfortunately, this transition
   has not been observed so far. The EPQFM approach is a microscopic one which takes care of LLL projection 
   from the very beginning. However, the charge sector was explicitly projected out,
   the connection and coupling between the charge sector
   which displays Fractional Quantum Hall effect and the spin sector
   which displays interlayer phase coherence was not obvious in the EPQFM approach.

   In \cite{mcf}, I used and extended both Mutual Composite Fermion (MCF) 
   and Composite Boson (CB) approaches to study both balanced and imbalanced BLQH and  made critical comparisons
   between the two approaches.
   I identified several serious problems with the MCF approach and then  developed
   a simple CB approach which naturally avoids all these problems suffered in the MCF approach.
   I found the CB approach is superior to MCF approach in the
   BLQH with broken symmetry. The functional form of
   the spin sector of the CB theory is the same as the EPQFM.
   It also has the advantage to treat charged excitations and the collective
   order parameter fluctuations in the pseudo-spin sector on the same footing,
   therefore can spell out the spin-charge connections explicitly. 
   From this spin-charge connection, we  are able to classify all the possible excitations
   in the BLQH in a systematic way.  The CB theory may also be applied to
   study the incoherent disordered insulating side and the quantum phase transitions
   between different ground states \cite{cbprl}.
    Using the CB approach, I also studied several interesting phenomena specific to
   im-balanced BLQH. Just like any Chern-Simon theory, the CB theory in BLQH has its own limitations:
   it is hard to incorporate the LLL projection ( see however \cite{shankar} ),
   some parameters can only be taken as phenomilogical parameters to be fitted
   into the microscopic LLL+HF calculations or experimental data. 
    It is an effective low energy theory, so special care is needed to
    capture some physics at microscopic length scales

  In this paper, we use both MCF approach and the CB approach developed in \cite{mcf}
  to study spin polarized Trilayer Quantum Hall (TLQH) systems at total filling factor
  $ \nu_{T} = 1 $. We also supplement the two approaches
  by wavefunction approach. TLQH is interesting from both experimental and theoretical sides.
  On the experimental side,
  TLQH systems have been fabricated in high mobility electron systems in the experimental group
  led by Shayegan \cite{tri}.
  On the theory side, because the three layers
  play the roles of three flavors. The excitons in BLQH is the pairing of particle in one layer
  and the hole in the other. This pairing structure in BLQH is similar to the Cooper pairing
  in BCS superconductors up to a particle-hole transformation. While the excitons in TLQH
  is not obvious, because so far there is no analog of three flavors superconductors yet.
  It would be interesting to understand the structures of the excitons and all the possible excitations
  in TLQH. As shown in this paper, due to the particular structure pattern of the excitons,
  there are two  {\em coupled} Goldstone modes in the TLQH
  ( see Eqn.\ref{tun3} ). In the presence of interlayer tunnelings ( with or without in-plane magnetic field ),
  the interference 
  between the two gapless modes may also lead to many new phenomena which can not be seen in BLQH.

  There are at least two well defined regimes for TLQH.
 (I) Interlayer coherent regime: when the distance $ d $ bewteen the two adjcent layers is sufficiently small,
     then all the three layers are strongly correlated.
  (II) Weakly-coupled regime: when $ d $  is sufficiently large, 
     then all the three layers are are weakly coupled. Depending on the distance $ d $, there could be many
      other possible regimes. 
  In this paper, we focus on regime (I) which is the interlayer coherent regime.
  The system in the interlayer coherent
  regime (I) was discussed in \cite{fer,hanna} in LLL+ HF approach, two
  Goldstone modes are found. Here, we study the regime (I) in detail from both MCF and CB approaches and
  also stress the broken symmetry state in terms of $ SU(3) $ algebra.

    Consider a tri-layer system with $ N_{1}, N_{2}, N_{3} $ electrons in layer 1 ( the bottom layer ),
  layer 2 ( the middle layer ) and layer 3 ( the top layer )
  respectively in the presence of magnetic field $ \vec{B} = \nabla \times \vec{A} $.
  We assume equal interlayer distance $ d $ between adjacent layers, the total filling factor is
  $ \nu_{T} =1 $ and the spin is completely polarized.
   The Hamiltonian $ H =  H_{0} + H_{int} $ is
\begin{eqnarray}
   H_{0} & = &  \int d^{2} x c^{\dagger}_{\alpha}(\vec{x})
   \frac{ (-i \hbar \vec{\nabla} + \frac{e}{c} \vec{A}(\vec{x}) )^{2} }
      {2 m }  c_{\alpha}(\vec{x})                \nonumber   \\
    H_{int} & = &  \frac{1}{2} \int d^{2} x d^{2} x^{\prime} \delta \rho_{\alpha} (\vec{x} )
               V_{\alpha \beta} (\vec{x}-\vec{x}^{\prime} )  \delta \rho_{\beta } ( \vec{x^{\prime}} )
\label{first}
\end{eqnarray}
  where electrons have bare mass $ m $ and carry charge $ - e $;
  $ c_{\alpha}, \delta \rho_{\alpha}(\vec{x}) = c^{\dagger}_{\alpha} (\vec{x})
  c_{\alpha} (\vec{x} ) -\bar{\rho}_{\alpha}, \alpha=1,2,3 $ are
  electron  operators and normal ordered electron densities on the three layers.
  The intralayer interactions are $ V_{11}=V_{22}= V_{33}=
  e^{2}/\epsilon r $, while interlayer interaction is $ V_{12}=V_{21}= V_{23}=V_{32}= e^{2}/ \epsilon
  \sqrt{ r^{2}+ d^{2} }, V_{13}=V_{31}= e^{2}/ \epsilon
  \sqrt{ r^{2}+ 4d^{2} } $ where $ \epsilon $ is the dielectric constant.

      The rest of the paper is organized as follows. In section II, we discuss MCF approach to
   TLQH and point out its weakness. In section III, we discuss CB approach and derive
   an effective action involving the charge and the two Goldstone modes.
   We also derive a dual action of the CB approach and make critical
   comparisons between this dual action and the MCF action. We classify all the possible excitations and
   discuss possible 2-body and 3-body bound states and associated Kosterlize-Thouless transitions.
   In section IV, we will outline several salient features of correlated interlayer
   tunnelings with or without in-plane magnetic field in TLQH.
   We reach conclusions in the final section. At appropriate places in the paper,
   we point out the limitations of the CB approach.

\section{ Mutual Composite Fermion approach:}

   We can extend the single-valued singular gauge transformation in \cite{mcf} to tri-layer system: 
\begin{equation}
  U= e^{ \frac{i}{2} \sum_{\alpha \beta} \int d^{2} x \int d^{2} x^{\prime}
  U_{\alpha \beta} \rho_{\alpha} (\vec{x} ) arg(\vec{x}-\vec{x}^{\prime} )
  \rho_{\beta} (\vec{x}^{\prime}) }
\label{sing}
\end{equation}
   where the $ 3 \times 3 $ symmetric matrix $ U $ and its inverse is given by:
\begin{equation}
   U_{t} =  \left ( \begin{array}{ccc}
      0 & 1 &  1   \\
      1 & 0 &  1   \\
      1 & 1 &  0   \\
   \end{array}   \right ) ~~~~~~~ U^{-1}_{t} = \frac{1}{2} \left ( \begin{array}{ccc}
      -1 & 1 &  1   \\
      1 & -1 &  1   \\
      1 & 1 &  -1   \\
   \end{array}   \right ) 
\label{inv}
\end{equation}

   The Hamiltonian Eqn.\ref{first} is transformed into:
\begin{equation}
   H_{0} = \int d^{2} x \psi^{\dagger}_{\alpha}(\vec{x}) \frac{ (-i \hbar \vec{\nabla} + \frac{e}{c} \vec{A}(\vec{x})
             - \hbar \vec{ a }_{\alpha}(\vec{x}) )^{2} }
      {2 m }  \psi_{\alpha}(\vec{x})
\label{second}
\end{equation}
  where the transformed fermion $ \psi_{\alpha}(\vec{x})  =  U c_{\alpha}(\vec{x}) U^{-1} $ is given by:
\begin{equation}
  \psi_{\alpha}(\vec{x})  =
   e ^{ \frac{i}{2} \int d^{2} x^{\prime} U_{\alpha \beta} [ arg( \vec{x} -\vec{x}^{\prime} ) +
   arg(\vec{x}^{\prime} -\vec{x} ) ] \rho_{\beta} ( \vec{x}^{\prime} ) } c_{\alpha}(\vec{x})
\end{equation}
  and the three Chern-Simon (CS) gauge fields $ a_{\alpha} $ in Eqn.\ref{second} satisfies:
  $ \nabla \cdot \vec{a}_{\alpha}=0, \nabla \times \vec{ a }_{\alpha} =
  2 \pi U_{\alpha \beta} \rho_{\beta}(\vec{x}) = 2 \pi
  [ \rho( \vec{x}) - \rho_{\alpha}(\vec{x}) ]  $ where
  $ \rho( \vec{x})= \sum_{\alpha} \rho_{\alpha}(\vec{x}) $
  is the total density of the system.

    In the following, we put $ \hbar=c=e= \epsilon = 1 $.
   At total filling factor $ \nu_{T}=1 $, $ \nabla \times \vec{A}
   = 2 \pi n $ where $ n=n_{1}+n_{2} +n_{3} $ is the total average electron density. By absorbing the average
   values of CS gauge fields $ \nabla \times < \vec{a}_{\alpha} > = 2 \pi (n-
   n_{\alpha} ) $ into the external gauge potential $ \vec{A}^{*}_{\alpha} =\vec{A}- < \vec{a}_{\alpha} > $,
   we have:
\begin{equation}
   H_{0} = \int d^{2} x \psi^{\dagger}_{\alpha}(\vec{x}) \frac{ (-i \vec{\nabla} + \vec{A}^{*}_{\alpha}(\vec{x})
             - \delta \vec{ a }_{\alpha}(\vec{x}) )^{2} }
      {2 m }  \psi_{\alpha}(\vec{x})
\label{original}
\end{equation}
  where $ \nabla \times \vec{A}^{*}_{\alpha} = 2 \pi n_{\alpha} $ and
  $  \nabla \times  \delta \vec{ a }_{\alpha} = 2 \pi U_{\alpha \beta}
  \delta \rho_{\beta}(\vec{x})= 2 \pi [ \delta \rho( \vec{x}) -
  \delta \rho_{\alpha}(\vec{x}) ] $ are the deviations from the corresponding
  average density.

   When $ d < d_{c}/2 $, the strong intra- and inter-layer interactions renormalize the bare mass into
   two different effective masses $ m^{*}_{1} = m^{*}_{3},  m^{*}_{2} $ \cite{shankar}.
   MCF in each layer feel effective magnetic
   field $ B^{*}_{\alpha} =\nabla \times \vec{A}^{*}_{\alpha} = 2 \pi n_{\alpha} $, therefore fill exactly
   one MCF Landau level. The energy gaps are simply the cyclotron gaps of the MCF
   Landau levels $ \omega^{*}_{c \alpha} = \frac{ B^{*}_{\alpha} }{ m^{*}_{\alpha} } $.

\underline{Transport properties:}
   Adding three different source gauge potentials $ \vec{A}^{s}_{ \alpha } $ 
   to the three layers in Eqn.\ref{original}, integrating out the MCF $ \psi_{\alpha} $
   first, then the mutual CS gauge fields, we can calculate the resistivity
   tensor by Kubo formula. We find that both Hall resistivity and
   two Hall drag resistivities are $ h/e^{2} $. This means if one drives current in one layer,
   the Hall resistivities in all the three layers are quantized at $ h/e^{2} $, while
   the longitudinal resistivities in all the three layers vanish. This is the hallmark
   of interlayer coherent quantum hall states.

   In the following, we confine to balanced case $ N_{1}=N_{2}=N_{3} = N/3 $, imbalanced case
   can be discussed along the similar line developed in \cite{mcf}.

\underline{ Fractional charges:}  Let's look at the charge of quasi-particles created by
   MCF field operators $ \psi^{\dagger}_{\alpha} ( \vec{x} ) $.
   If we insert one electron at one of the layers, say layer 1, from the singular gauge transformation in Eqn.\ref{sing},
   we can see this is equivalent to insert one MCF in layer 1, at the same time, insert one flux quantum
   at layer 2 and another one at layer 3 directly above the position where we inserted the electron.
   The inserted flux quantum at layer 2 and layer 3 are in the opposite direction to the external
   magnetic field, therefore promote one MCF in layer 2 and one in layer 3 to the second MCF Landau level
   in layer 2 and layer 3 respectively. Because inserting one electron leads to three MCFs in each layer,
   we conclude the charge of each MCF is -1/3.
    The above argument gives the correct {\em total} fractional charges of MCF, but
    it can not determine the {\em relative} charge  distributions among the three layers.
    At mean field level, the energies of all the possible relative charge differences are degenerate.
    The lowest energy configuration can only be determined by fluctuations.

\underline{ Left- and right-moving  gapless modes:} 
   $ \delta \rho_{\alpha}(\vec{x}) $ can be expressed in terms of the CS gauge fields $
   \delta \rho_{\alpha}(\vec{x}) = [U_{t}^{-1}]_{\alpha \beta}
   \frac{ \delta \vec{ a }_{\beta} }{ 2 \pi} $ where
   $ U_{t}^{-1} $ is given in Eqn.\ref{inv}.
    This constraint can be imposed by Lagrangian multipliers $ a^{0}_{\alpha} $
   which plays the role of time components of CS gauge fields.
   Integrating out MCF $ \psi_{1}, \psi_{2},  \psi_{3} $ to one-loop and carefully expanding the interlayer Coulomb
   interaction to the necessary order in the long-wavelength limit leads to an effective action for
   the three gauge fields.
\begin{eqnarray}
  {\cal L} & = & \frac{ i q}{ 4 \pi}  a^{0}_{c} a^{t}_{c} +  \frac{q}{ 16 \pi } (  a^{t}_{c} )^{2}
                         \nonumber   \\
         & +  & \frac{ \epsilon_{c} }{6} q^{2} ( a^{0}_{c} )^{2} 
         + \frac{1}{6}  [ \epsilon_{c} \omega^{2} + ( \chi_{c}
             -\frac{ d }{ 3 \pi} ) q^{2} ] ( a^{t}_{c} )^{2} 
                        \nonumber  \\
         & +  & \frac{ \epsilon_{l} }{4}  q^{2} ( a^{0}_{l} )^{2} 
         + \frac{1}{4} [ \epsilon_{l} \omega^{2} + ( \chi_{l} +
          \frac{ d}{ \pi} ) q^{2} ] ( a^{t}_{l} )^{2} 
                        \nonumber  \\
         & +  & \frac{ \epsilon_{r} }{12}  q^{2} ( a^{0}_{r} )^{2} 
         + \frac{1}{12} [ \epsilon_{r} \omega^{2} +
      ( \chi_{r} + \frac{ d}{ 3 \pi} ) q^{2} ] ( a^{t}_{r} )^{2} 
                        \nonumber  \\
         & +  & \frac{ \delta \epsilon }{9} q^{2}  a^{0}_{c} a^{0}_{r}
           + \frac{1}{9}[ \delta \epsilon \omega^{2} + ( \delta \chi
              + \frac{d}{4 \pi} ) q^{2} ] a^{t}_{c} a^{t}_{r} + \cdots 
\label{ngm}
\end{eqnarray}
   where $ \vec{a}_{s}= \vec{a}_{1} + \vec{a}_{2} + \vec{a}_{3},
   \vec{a}_{l}= \vec{a}_{1} - \vec{a}_{3},
   \vec{a}_{r}= \vec{a}_{1} + \vec{a}_{3} - 2 \vec{a}_{2} $ which stand for center of mass,
   left and right channels respectively \cite{note}. $ \epsilon_{c}=
   ( 2 \epsilon_{1} + \epsilon_{2} )/3, \chi_{c}= (2 \chi_{1} + \chi_{2})/3,
   \epsilon_{l}= \epsilon_{1}, \chi_{l}= \chi_{1},
   \epsilon_{r}= ( \epsilon_{1} + 2 \epsilon_{2} )/3, \chi_{r}= ( \chi_{1} + 2 \chi_{2})/3,
    \delta \epsilon = \epsilon_{1} - \epsilon_{2}, \delta \chi= \chi_{1} - \chi_{2} $.
   The dielectric constants
   $ \epsilon_{\alpha} = \frac{ m^{*}_{\alpha} }{ 2 \pi B^{*}_{\alpha} } $ and the
   susceptibilities $ \chi_{\alpha}= \frac{1}{ 2 \pi m^{*}_{\alpha} } $
   were calculated in single layer system in \cite{frad}.
   $ \cdots $ are higher gradient terms and
   $ a^{t}_{\alpha} $ is the transverse component of gauge fields in Coulomb gauge
   $ \nabla \cdot \vec{a}_{\alpha} =0 $.
  The system is invariant under the $ Z_{2} $ symmetry which exchanges
   layer 1 and 3, namely $ a_{c} \rightarrow a_{c}, a_{l} \rightarrow -a_{l}, a_{r} \rightarrow a_{r} $.
   This symmetry dictates that only $ a_{c} a_{c}, a_{c} a_{r},
   a_{r} a_{r}, a_{l} a_{l} $ can appear in
   the Maxwell terms. The last two terms in Eqn. \ref{ngm} which is the coupling between
   $ a_{c} $ and $ a_{r} $ can be shown to be
   irrelevant in the low energy limit.

  More intuitive choices are $ a_{u} = a_{1}-a_{2}, a_{d} = a_{2}-a_{3} $ which is related to
  $ a_{l}, a_{r} $ by $ a_{l}= a_{u} + a_{d}, a_{r}= a_{u} - a_{d} $. But $ a_{u} $ and
  $ a_{d} $ are coupled, the two normal modes are $ a_{l}, a_{r} $
  instead of $ a_{u}, a_{d} $.  
  As expected, there is a Chern-Simon term for the center of mass
  gauge field $ \vec{a}_{s} $,  while there are two Maxwell terms for the
  left and right channel gauge fields $ \vec{a}_{l}, \vec{a}_{r} $ which are gapless.
  The two gapless modes stand for the relative charge density fluctuations among the three layers.
  After putting back $ \hbar, c , e, \epsilon $, we find the two spin-wave velocities in terms of
  experimentally measurable parameters:
\begin{eqnarray}
  v^{2}_{l} & = & \frac{ 3 ( \omega^{*}_{c} )^{2} }{ 2 \pi n } +
   (  \frac{ 2 \alpha c}{\epsilon} ) ( \frac{d}{l} ) \frac{ \omega^{*}_{c} }{ \sqrt{ 2 \pi n} } 
                    \nonumber \\
  v^{2}_{r} & = & \frac{ 3 ( \omega^{*}_{c} )^{2} }{ 2 \pi n }
  \frac{1+ 2 \gamma^{-1} }{ 1+ 2 \gamma } +
   (  \frac{ 2 \alpha c}{  \epsilon} ) ( \frac{d}{l} )
   \frac{ \omega^{*}_{c} }{ \sqrt{ 2 \pi n} } \frac{1}{ 1+ 2 \gamma } 
\label{v}
\end{eqnarray}
   where we set $ \omega^{*}_{c}= \omega^{*}_{c1} $ and $ n $ is
   the total density, $ l $ is the magnetic length, $ \gamma $ is the ratio
   of the two effective masses $ \gamma= m^{*}_{2}/m^{*}_{1} $
   and $ \alpha \sim 1/137 $ is the fine structure constant.

   As in BLQH \cite{mcf}, we expect the second term dominates the first,
   then $ v_{l}/v_{r} \sim \sqrt{ 1 + 2 \gamma } $. namely, the ratio of the two
   velocities is determined by the ratio of the two effective masses.
   If we take the effective masses to be the band mass, then $  v_{l}/v_{r} \sim \sqrt{3} $.


\underline{ Topological excitations:} As discussed in the previous paragraph, the MCF carries charge $ 1/3 $,
   but their relative charge distributions among the three layers are undetermined.
   They can be characterized 
   by their $ ( a_{c}, a_{l}, a_{r} ) $ charges $ ( \pm 1/3, q_{l}, q_{r} ) $ with
   $ q_{l}, q_{r}= 0, \pm 1, \cdots $.
   Exchanging $ a_{c} $ leads to $ 1/r $ interaction between two MCF, while
   exchanging $ a_{l}, a_{r} $ leads to logarithmic interactions which may lead to a 2-body bound state 
   between two MCF with opposite $ ( q_{l}, q_{r} ) $.
   The energy of this bound state with length $ L $
   is $ \Delta_{+} + \Delta_{-} + q^{2}_{c} \frac{ e^{2} }{ L }
    - ( q_{l1} q_{l2} + \frac{ q_{r1} q_{r2} }{1 + 2 \gamma } )
   \hbar \omega^{*}_{c} \ln L/l $
   where $ \Delta_{+}, \Delta_{-} $ are the core energies of QH and QP respectively.
   There are also possible 3-body bound states of three MCF with $ \sum_{i} q_{li}=\sum_{i} q_{ri} =0 $.

   Unfortunately, the gluing conditions ( or selection rules ) of $ (q_{c}, q_{l}, q_{r} ) $
   for realizable physical excitations  are not clear. This charge $ q_{c} $ and two spin $ ( q_{l}, q_{r} ) $
   connections can only be easily established from CB approach presented in the next section.
   So we defer the discussion on the classification of all the possible excitations to the next section.

\underline{ Problems with MCF approach:}

   All the criticisms on MCF approach to BLQH also hold to TLQH. In the following,
   we list just  ones which are the most
   relevant to this paper.

 (a)  It is easy to see that the spin wave dispersion in Eqn.\ref{v} remains
   linear $ \omega \sim v k $ even in the
   $ d \rightarrow 0 $ limit. This contradicts with the well established fact that in the
   $ d \rightarrow 0 $ limit, the linear dispersion relation
   will be replaced by quadratic $ SU(3) $ Ferromagnetic spin-wave
   dispersion relation $ \omega \sim  k^{2} $ due
   to the enlarged $ SU(3) $ symmetry at $ d \rightarrow 0 $.

 (b) The broken symmetry in the ground state is not obvious
    without resorting to the $ (111,111) $ wavefunction Eqn.\ref{ground}.
    The origin of the gapless mode is not clear.

 (c)  The physical meaning of $ q_{l}, q_{r} $ is not clear. That they have to be integers
    was put in by hand in an ad hoc way.

 (d) The spin-charge connections in $ ( q_{c}, q_{l}, q_{r} ) $ can not be extracted.

 (e) Even in the balanced case, there are two MCF cyclotron gaps $ \hbar \omega^{*}_{c \alpha}, \alpha=1, 2 $
    at mean field theory. However, there is only one charge gap in the system.
    It is not known how to reconcile this discrepancy within MCF approach.

 (f) In the presence of interlayer tunneling, it is not known how to derive the tunneling term 
     in a straightforward way ( see section IV).

   In the following, we will show that the alternative CB approach not only achieve
   all the results, but also can get rid of all
   these drawbacks.

\section{Composite Boson approach }
 
     Composite boson approach was found to be much more effective than the MCF approach
   in BLQH. We expect it remains true in TLQH. In this section, we apply CB approach
   to study TLQH and find that it indeed avoids all the problems suffered in MCF approach
   listed in the last section.
   Instead of integrating out the charge degree of freedoms which was done in the previous
   CB approach, we keep one charge sector
   and  two spin sectors on the same footing and explicitly stress the spin and charge
   connections. We classify all the possible excitations, bound states and associated
   KT transitions.

      By defining the center of mass, left and right channels densities as:
\begin{eqnarray}
   \delta \rho_{c} & =  & \delta \rho_{1} +  \delta \rho_{2} + \delta \rho_{3}
                                         \nonumber  \\
   \delta \rho_{l} & = &  \delta \rho_{1} - \delta \rho_{3}
                                         \nonumber  \\
   \delta \rho_{r} & =  & \delta \rho_{1} -  2 \delta \rho_{2} + \delta \rho_{3}
\label{3d}
\end{eqnarray}
     We can rewrite $ H_{int} $ in Eqn.\ref{first} as:
\begin{eqnarray}
   H_{int} & = & \frac{1}{2} \delta \rho_{1} U  \delta \rho_{1} +  \frac{1}{2} \delta \rho_{2} U  \delta \rho_{2}
   + \frac{1}{2} \delta \rho_{3} U  \delta \rho_{3}   \nonumber   \\
    & + &  \delta \rho_{1}  V  \delta \rho_{2}
   + \delta \rho_{2}  V  \delta \rho_{3}
   + \delta \rho_{1}  W  \delta \rho_{3}
                    \nonumber  \\
   & =  & \frac{1}{2} \delta \rho_{c} V_{c} \delta \rho_{c}
   + \frac{1}{2}  \delta \rho_{l} V_{l} \delta \rho_{l}
   + \frac{1}{2}  \delta \rho_{r} V_{r} \delta \rho_{r}
   +  \delta \rho_{c} V_{cr} \delta \rho_{r}
\end{eqnarray}
     where $ U=V_{11}, V=V_{12}, W=V_{13} $ and $ V_{c}=\frac{ U}{ 3} + \frac{4 V}{ 9 } + \frac{ 2 W}{ 9},
       V_{l}= \frac{1}{2} ( U-W), V_{r}= \frac{1}{3}( \frac{U}{2} -\frac{ 2 V}{3} + \frac{ w}{6} ),
       V_{cr}= - \frac{1}{9} ( V-W ) $. The stability of the system dictates $ V_{c} V_{r} - V^{2}_{cr} > 0 $.
      In the long wavelength limit $ q d \ll 1 $, we only keep
      the leading terms: $ V_{c} = \frac{ 2 \pi}{q},
       V_{l}= 2 \pi d, V_{r}= \frac{ 2 \pi d}{9}, V_{cr} = - \frac{ 2 \pi d }{ 9 } $.

     Performing a singular gauge transformation:
\begin{equation}
  \phi_{a}(\vec{x}) = e ^{ i \int d^{2} x^{\prime} arg(\vec{x}-\vec{x}^{\prime} ) 
  \rho ( \vec{x}^{\prime} ) } c_{a}( \vec{x})
\label{singb}
\end{equation}
     where $ \rho ( \vec{x} ) = c^{\dagger}_{1}( \vec{x} ) c_{1}( \vec{x} ) +
     c^{\dagger}_{2}( \vec{x} ) c_{2}( \vec{x} )  +
     c^{\dagger}_{3}( \vec{x} ) c_{3}( \vec{x} ) $ is the total density of the tri-layer system.

     After absorbing the external gauge potential $ \vec{A} $ into the Chern-Simon gauge  potential
  $ \vec{a} $, we can transform the Hamiltonian Eqn.\ref{first} into the Lagrangian in the Coulomb gauge:
\begin{eqnarray}
   {\cal L} & = & \phi^{\dagger}_{a}( \partial_{\tau}- i a_{0} ) \phi_{a}
    + \phi^{\dagger}_{a}(\vec{x}) \frac{ (-i \hbar \vec{\nabla} 
             - \hbar \vec{ a }(\vec{x}) )^{2} }{2 m }  \phi_{a}(\vec{x})
                      \nonumber  \\     
     & + & i a_{0} \bar{\rho}
     + \frac{1}{2} \int d^{2} x^{\prime} \delta \rho_{c} (\vec{x} ) 
               V_{c} (\vec{x}-\vec{x}^{\prime} )  \delta \rho_{c} ( \vec{x^{\prime}} )
                           \nonumber  \\
    & + & \frac{1}{2} \int d^{2} x^{\prime} \delta \rho_{l} (\vec{x} ) 
               V_{l} (\vec{x}-\vec{x}^{\prime} )  \delta \rho_{l} ( \vec{x^{\prime}} )
                         \nonumber  \\
    & +  &  \frac{1}{2} \int d^{2} x^{\prime} \delta \rho_{r} (\vec{x} ) 
               V_{r} (\vec{x}-\vec{x}^{\prime} )  \delta \rho_{r} ( \vec{x^{\prime}} )
                    \nonumber   \\
     & +  &   \int d^{2} x^{\prime} \delta \rho_{c} (\vec{x} ) 
               V_{cr} (\vec{x}-\vec{x}^{\prime} )  \delta \rho_{r} ( \vec{x^{\prime}} )
    -\frac{ i }{ 2 \pi} a_{0} ( \nabla \times \vec{a} ) 
\label{boson}
\end{eqnarray}

   In the Coulomb gauge, integrating out $ a_{0} $ leads to the constraint: $ \nabla \times \vec{a}
   = 2 \pi ( \phi^{\dagger}_{a} \phi_{a} - \bar{\rho} ) $.

   It can be shown that $ \phi_{a}(\vec{x}) $ satisfies all the boson commutation relations.
   We write the three bosons in terms of magnitude and phase
\begin{equation}
  \phi_{a}= \sqrt{ \bar{\rho}_{a} + \delta \rho_{a} } e^{i \theta_{a} }
\label{mag}
\end{equation}
   The boson commutation relations imply that
   $ [ \delta \rho_{a} ( \vec{x} ), \phi_{b}( \vec{x} ) ] =
     i \hbar \delta_{ab} \delta( \vec{x}-\vec{x}^{\prime} ) $.
   For simplicity, in this paper, we only consider the balanced case, so we set
   $ \bar{\rho}_{a} = \bar{\rho}  $. The imbalanced
   case can be discussed along similar lines developed in \cite{mcf}.

    We define the center of mass, left-moving and right-moving angles
    which are conjugate angle variables to the three densities defined in Eqn.\ref{3d}
\begin{eqnarray}
    \theta_{c} & =  & \theta_{1} +  \theta_{2} + \theta_{3}
                                         \nonumber  \\
   \theta_{l} & = &  \theta_{1} - \theta_{3}
                                         \nonumber  \\
   \theta_{r} & =  & \theta_{1} -  2 \theta_{2} + \theta_{3}
\label{an}
\end{eqnarray}
    which satisfy the commutation relations
    $ [ \delta \rho_{\alpha} ( \vec{x} ), \theta_{\beta}( \vec{x}^{\prime} ) ] =
    A_{\alpha} i \hbar \delta_{\alpha \beta} \delta( \vec{x}-\vec{x}^{\prime} ) $ where $ A=3, 2, 6 $
    for $ \alpha = c, l, r $.

    Substituting Eqn\ref{mag} into Eqn.\ref{boson},
    neglecting the magnitude fluctuations in the spatial
    gradient term which were shown to be irrelevant in BLQH in \cite{mcf}
    and rewriting the spatial gradient term
    in terms of the three angles in Eqn.\ref{an}, we find:
\begin{eqnarray}
  {\cal L}  &  =  & i \delta \rho_{c} ( \frac{1}{3} \partial_{\tau} \theta_{c}-  a_{0} ) + 
          \frac{ \bar{\rho} }{2m} [ \frac{1}{3} \nabla \theta_{c} 
          - \vec{a} ]^{2}
     +  \frac{1}{2} \delta \rho_{c} V_{c} (\vec{q} )  \delta \rho_{c}
                       \nonumber   \\
    & + & \frac{i}{2} \delta \rho_{l}  \partial_{\tau} \theta_{l} + 
          \frac{ \bar{\rho} }{12m} ( \nabla \theta_{l} )^{2}
          + \frac{1}{2} \delta \rho_{l} V_{l} (\vec{q} )  \delta \rho_{l}
                      \nonumber  \\
    & + & \frac{i}{6} \delta \rho_{r}  \partial_{\tau} \theta_{r} + 
          \frac{ \bar{\rho} }{36m} ( \nabla \theta_{r} )^{2}
          + \frac{1}{2} \delta \rho_{r} V_{r} (\vec{q} )  \delta \rho_{r}
                        \nonumber  \\
          & + & \delta \rho_{c} V_{cr} (\vec{q} )  \delta \rho_{r}
          - \frac{ i }{ 2 \pi} a_{0} ( \nabla \times \vec{a} ) 
\label{main}
\end{eqnarray}
     
  In the balanced case, the symmetry is $ U(1)_{c} \times U(1)_{l} \times U(1)_{r} \times Z_{2} $
  where the first is a local gauge symmetry, the second and the third are global symmetries and
  the global $ Z_{2} $ symmetry is the exchange symmetry between layer 1 and layer 3.

{\sl (a) Off-diagonal Algebraic order in the charge sector: }

    At temperatures much lower than the vortex excitation energy, we can neglect vortex configurations
  in Eqn.\ref{main} and only consider the low energy spin-wave excitation.
  The charge sector
  ( $ \theta_{c} $ mode ) and the two spin sectors ( $ \theta_{l},  \theta_{r} $ modes ) are essentially decoupled.
  It can be shown that the second to the last term 
          $ \delta \rho_{c} V_{cr} (\vec{q} )  \delta \rho_{r} $ in Eqn.\ref{main}
  which is the coupling between the charge sector and right-moving sector
     is irrelevant in the long wave-length limit. Therefore,
    the charge sector is essentially the same as the CSGL action in BLQH.
    Using the constraint $ a_{t} = \frac{ 2 \pi \delta \rho_{c} }{ q } $, neglecting
   vortex  excitations in the ground state and integrating out $ \delta \rho_{c} $ leads to
   the effective action of $ \theta_{c} $:
\begin{equation}
    {\cal L}_{c} = \frac{1}{2} \times ( \frac{1}{3} )^{2} \theta_{c} (- \vec{q}, - \omega )
       [ \frac{ \omega^{2} + \omega^{2}_{ \vec{q} } }{ V_{c}( q) + \frac{ 4 \pi^{2} \bar{\rho} }{m}
     \frac{1}{ q^{2} } } ] \theta_{c} ( \vec{q},  \omega )
\label{cy}
\end{equation}
    where $ \omega^{2}_{ \vec{q} } = \omega^{2}_{c} + \frac{ \bar{\rho} }{m} q^{2} V_{c}( q) $
     and $ \omega_{c} = \frac{ 2 \pi \bar{\rho} }{ m } $ is the cyclotron frequency.

    From  Eqn.\ref{cy}, we can find the equal time correlator of $ \theta_{c} $:
\begin{eqnarray}
     < \theta_{c}( - \vec{q} ) \theta_{c} ( \vec{q} ) > & = &
      \int^{\infty}_{-\infty}
      \frac{ d \omega }{ 2 \pi}
      < \theta_{c} (- \vec{q}, - \omega ) \theta_{c} ( \vec{q}, \omega ) >
                                 \nonumber    \\
     & =  & 3 \times \frac{2 \pi}{ q^{2} }  + O( \frac{1}{ q } ) 
\label{eq}
\end{eqnarray}
     which leads to the algebraic order:
\begin{equation}
     < e^{ i ( \theta_{c}(\vec{x}) - \theta_{c}( \vec{y} ) ) } > = \frac{1}{ | x-y |^{3} }
\label{exp}
\end{equation}

  We could define $ \tilde{ \theta}_{c}  = ( \theta_{1} + \theta_{2} + \theta_{3} )/3 = \theta_{c}/3 $, then the exponent
  will be $ 1/3 $. But when we consider topological vortex excitations, then $ e^{ i \tilde{ \theta }_{c} } $ may not be
  single valued. Therefore, $ \theta_{c} $ is more fundamental than $ \tilde{\theta}_{c} $.

{\sl (b) Spin wave excitations: }

  While in the spin sector, there are two
  neutral gapless modes: left-moving mode and right-moving mode \cite{note}.
  Integrating out $ \delta \rho_{l} $ leads to
\begin{equation}
    {\cal L}_{l} = \frac{1}{ 2 V_{l}( \vec{q} ) } ( \frac{1}{2} \partial_{\tau} \theta_{l} )^{2} + 
     \frac{ \bar{\rho} }{12m} (  \nabla \theta_{l} )^{2}
\end{equation}
    where the dispersion relation of spin wave can be extracted:
\begin{equation}
     \omega^{2} = [ \frac{ 2 \bar{\rho} }{ 3 m}  V_{l}( \vec{q} ) ] q^{2} = v^{2}_{l}  q^{2}
\end{equation}
     In the long wave-length limit $ qd \ll 1 $, the bare spin wave velocity is:  
\begin{equation}
     v^{2}_{l} = \frac{ \bar{\rho} }{m} \frac{ 4 \pi e^{2} }{ 3 \epsilon}  d  = \frac{ 2 e^{2} }{ 3 m \epsilon}
       \sqrt{ 2 \pi \bar{\rho} } ( \frac{ d}{l}  )
\label{vl}
\end{equation}

  Integrating out $ \delta \rho_{r} $ leads to
\begin{equation}
    {\cal L}_{r} = \frac{1}{ 2 V_{r}( \vec{q} ) } ( \frac{1}{6} \partial_{\tau} \theta_{r} )^{2} + 
     \frac{ \bar{\rho} }{36m} ( \nabla \theta_{r} )^{2}
\label{r1}
\end{equation}
    where the dispersion relation of spin wave can be extracted:
\begin{equation}
     \omega^{2} = [ \frac{ 2 \bar{\rho} }{  m}  V_{r}( \vec{q} ) ] q^{2} = v^{2}_{r}  q^{2}
\label{r2}
\end{equation}
     In the long wave-length limit $ qd \ll 1 $, the bare spin wave velocity is:  
\begin{equation}
     v^{2}_{r} = \frac{ \bar{\rho} }{m} \frac{ 4 \pi e^{2} }{ 9 \epsilon}  d  = v^{2}_{l}/3
\label{vr}
\end{equation}

  Both spin wave velocities should increase as the square root of the separation $ d $ when $ d < d_{c}/2 $,
  their ratio $  v_{l}/v_{r} \sim \sqrt{3} $.
  At $ d=0 $, $ v_{l} = v_{r} =0 $. This is expected, because at $ d=0 $ the
  $ U(1)_{l} \times U(1)_{r} \times Z_{2} $ symmetry is enlarged to
  $ SU(3)_{G} $, the spin wave of $ SU(3)  $ isotropic ferromagnet $ \omega \sim k^{2} $.

  We have also treated the coupling term between the charge and right-moving sector $ 
  \delta \rho_{c} V_{cr} (\vec{q} )  \delta \rho_{r} $ in Eqn.\ref{main} exactly
  and found that in the $ \vec{q}, \omega \rightarrow 0 $ limit, the mixed propagator
  takes the form:
\begin{equation}
  < \theta_{c}(-\vec{q}, - \omega) \theta_{r}(\vec{q},  \omega) >
    =\frac{ \frac{9}{2} ( \frac{ m }{ \pi \bar{ \rho} } )^{2} V_{cr} \omega^{2} } { \omega^{2} + v^{2}_{r} q^{2} }
    \sim \omega^{2} < \theta_{r} \theta_{r} >
\end{equation}
    which is dictated by $ < \theta_{r} \theta_{r} > $ instead of $ < \theta_{c} \theta_{c} > $.
    Because of the $ \omega^{2} $ prefactor, we can ignore $ < \theta_{c} \theta_{r} > $ 
    in the low energy limit.

  We also found that Eqns.\ref{cy}, \ref{eq}, \ref{exp} and Eqns.\ref{r1},\ref{r2},\ref{vr} remain true in
  the long-wavelength and low energy limit, therefore confirm that 
  $ \delta \rho_{c} V_{cr} (\vec{q} )  \delta \rho_{r} $ is indeed irrelevant in the limit.

   As shown in BLQH in \cite{mcf}, the functional form of the spin sector in the CB theory is the same
   as EPQFM. Although there are no detailed microscopic
   calculations in TLQH yet except the preliminary work in \cite{fer,hanna}, from the insights gained in BLQH,
   we can claim that the functional form of the spin sector in Eqn.\ref{main}
   ( or Eqn.\ref{tun3} ) achieved in the CB theory is correct. Again, 
   it is hard to incorporate the LLL projection in this CB approach,
   So some parameters can not be evaluated within this approach.
   For example, the two spin stiffnesses
   for the left and right-moving sectors in Eqn.\ref{main} ( or or Eqn.\ref{tun3} )
   should be completely determined by the Coulomb interactions instead of
   being dependent of the band mass, they  may be estimated by the microscopic
   LLL+ HF approach.  However, the ratio of the two velocities in Eqn.\ref{vr} is independent of
   the band mass, it is interesting to see if this ratio $ \sqrt{3} $
   remains correct in the microscopic LLL+HF estimation.
   Just like in BLQH, we can simply take $ \rho_{sl}, V_{l} $ and $ \rho_{sr}, V_{r} $ in Eqn.\ref{main} as
   phenomilogical parameters to be fitted into the LLL  + HF estimations.
   However, just like in BLQH \cite{moon,mcf,wave}, because the qualitative correct ground state
   wavefunction in TLQH is still unknown, it is still difficult to estimate these parameters with controlled
   accuracy even in the LLL+HF approach. We can simply fit these parameters into experimental data. 

{\sl (c) Dual action: }

  Just like in BLQH discussed in \cite{mcf}, we can perform a duality transformation on Eqn.\ref{main}
  to obtain a dual action in terms of three vortex currents $ J^{v;c,l,r}_{\mu} = \frac{1}{ 2 \pi} 
  \epsilon_{\mu \nu \lambda} \partial_{\nu} \partial_{\lambda} \theta_{c,l,r} $
  and three dual gauge fields $ b^{c}_{\mu}, b^{l}_{\mu}, b^{r}_{\mu} $:
\begin{eqnarray}
   {\cal L}_{d} & = & - i \pi b^{c}_{\mu} \epsilon_{\mu \nu \lambda} \partial_{\nu} b^{c}_{\lambda}
   - i A^{c}_{ s \mu} \epsilon_{\mu \nu \lambda} \partial_{\nu} b^{c}_{\lambda}
    + i \frac{ 2 \pi}{3} b^{c}_{\mu} J^{vc}_{\mu}
                      \nonumber   \\
   & + & \frac{ m }{ 2 \bar{\rho} f } ( \partial_{\alpha} b^{c}_{0}
      - \partial_{0} b^{c}_{\alpha} )^{2} + \frac{1}{2} ( \nabla \times \vec{b}^{c} ) V_{c} ( \vec{q})
      ( \nabla \times \vec{b}^{c} )
                            \nonumber   \\
   & - &  i A^{l}_{ s \mu} \epsilon_{\mu \nu \lambda} \partial_{\nu} b^{l}_{\lambda}
    + i \pi b^{l}_{\mu} J^{vl}_{\mu}
                      \nonumber   \\
   & + &  \frac{ 3 m }{ 4 \bar{\rho}  } ( \partial_{\alpha} b^{l}_{0}
      - \partial_{0} b^{l}_{\alpha} )^{2} + \frac{1}{2} ( \nabla \times \vec{b}^{l} ) V_{l} ( \vec{q})
      ( \nabla \times \vec{b}^{l} )
                      \nonumber   \\
   & - &  i A^{r}_{ s \mu} \epsilon_{\mu \nu \lambda} \partial_{\nu} b^{r}_{\lambda}
    + i \frac{\pi}{3} b^{r}_{\mu} J^{vr}_{\mu}
                      \nonumber   \\
   & + &  \frac{  m }{ 4 \bar{\rho}  } ( \partial_{\alpha} b^{r}_{0}
      - \partial_{0} b^{r}_{\alpha} )^{2} + \frac{1}{2} ( \nabla \times \vec{b}^{r} ) V_{r} ( \vec{q})
      ( \nabla \times \vec{b}^{r} )
                      \nonumber   \\
    & + & ( \nabla \times \vec{b}^{c} ) V_{cr} ( \vec{q}) ( \nabla \times \vec{b}^{r} )
\label{dual}
\end{eqnarray}
   where the three source gauge fields are $ A^{c}_{s \mu} = A_{1 s \mu} + A_{2 s \mu} + A_{ 3 s \mu},
   A^{l}_{s \mu} = A_{1 s \mu} - A_{ 3 s \mu}, A^{r}_{s \mu} = A_{1 s \mu} - 2 A_{2 s \mu} + A_{ 3 s \mu} $.

  We can compare this dual action
  derived from CB approach with Eqn.\ref{ngm} derived in MCF approach. Just like in BLQH, the three topological
  vortex currents are missing in Eqn.\ref{ngm}, while $ \chi_{c}, \chi_{l}, \chi_{r}, \delta \chi $ terms are
  extra spurious terms which break $ SU(3) $ symmetry even in the $ d \rightarrow 0 $ limit. If we drop all the $ \chi $
  terms, use bare mass and add the three vortex currents terms in Eqn.\ref{ngm}, the resulting effective action
  will be identical to the dual action Eqn.\ref{dual}. Neglecting the vortex currents, we can identify
  the two spin wave velocities easily
  from the dual action which are the same as found previously from Eqn.\ref{main}. It is also easy to show that
  the last term in Eqn.\ref{dual} which is the coupling between the charge and right-moving sector is irrelevant
  in the low energy limit.  We conclude that Eqns.\ref{main} and its dual action Eqn. \ref{dual}
  are the correct and complete effective actions.  

{\sl (d) Topological excitations:}

  Any topological excitations are characterized by three winding numbers
  $ \Delta \theta_{1}= 2 \pi m_{1}, \Delta \theta_{2}= 2 \pi m_{2}, \Delta \theta_{3}=
   2 \pi m_{3} $, or equivalently,
  $ \Delta \theta_{c}= 2 \pi ( m_{1} + m_{2} + m_{3} ) = 2 \pi m_{c}, 
    \Delta \theta_{l}= 2 \pi ( m_{1} - m_{3} ) = 2 \pi m_{l},
    \Delta \theta_{r}= 2 \pi ( m_{1} - 2 m_{2} + m_{3} ) = 2 \pi m_{r} $.
  It is important to stress that the three fundamental angles
  are $ \theta_{1}, \theta_{2}, \theta_{3} $ instead of
  $ \theta_{c}, \theta_{l}, \theta_{r} $. Therefore, $ m_{1}, m_{2}, m_{3} $ are three
  independent integers, while $ m_{c}, m_{l}, m_{r} $ are not.

  There are following 6 kinds of fundamental topological excitations: $ \Delta \phi_{1} = \pm 2 \pi $
  or  $ \Delta \phi_{2} = \pm 2 \pi $ or  $ \Delta \phi_{3} = \pm 2 \pi $,
  namely $ ( m_{1}, m_{2}, m_{3} ) = ( \pm 1, 0, 0 ), ( 0, \pm 1, 0 ), ( 0, 0, \pm 1 ) $.
  They correspond to inserting one flux quantum in layer 1 or 2 or 3
  in the same or opposite direction as the external magnetic field.
   Let's classify the topological excitations in terms of $ (q, m_{l}, m_{r} ) $
  where  $ q $ is the fractional charge of the topological excitations in the following table.

\vspace{0.25cm}
\begin{tabular}{ |c|c|c|c| }
             &  $ ( \pm 1, 0, 0 ) $  &  $ ( 0, \pm 1, 0) $  &  $ (0,0, \pm 1) $             \\  \hline
  $ m_{c} $  &  $  \pm 1 $        &   $ \pm 1 $    &  $ \pm 1 $               \\   \hline
  $ m_{l} $  &  $  \pm 1 $        &   $  0  $    &  $ \mp 1 $               \\   \hline
  $ m_{r} $  &  $  \pm 1 $        &   $ \mp2 $    &  $ \pm 1 $              \\   \hline
   $ q    $  &  $ \pm 1/3 $        &   $ \pm 1/3 $    &  $ \pm 1/3 $        
\end{tabular}
\par
\vspace{0.25cm}
{\footnotesize  {Table 1: The fractional charge in balanced TLQH } }
\vspace{0.25cm}

  The fractional charges in the table were determined from the constraint  
  $ \nabla \times \vec{a} = 2 \pi \delta \rho_{c} $ and the finiteness
  of the energy in the charge sector:
\begin{equation}
  q= \frac{1}{ 2 \pi} \oint  \vec{a} \cdot d \vec{l}
  = \frac{1}{ 2 \pi} \times \frac{1}{3} \oint \nabla \theta_{c} \cdot d \vec{l} =
    \frac{1}{3} m_{c}
\label{charge}
\end{equation}

    There are the following two possible bound states:

 (1) Three Charge neutral two-body bound states:
\begin{itemize}
\item
  Layer 1:  $  ( \pm 1/3, \pm 1,  \pm 1 ) $ with energy $ E_{1}= E_{c+} + E_{c-} - \frac{ e^{2} }{ 9 R }
        + 2 \pi ( \rho_{sl} +  \rho_{sr} ) \ln \frac{R}{R_{c}} $ when $ R \gg R_{c} \gg l $.
\item
  Layer 2: $  ( \pm 1/3, 0,  \mp 2 ) $ with energy $ E_{2}= E_{c+} + E_{c-} - \frac{ e^{2} }{ 9 R }
        + 2 \pi ( 4 \rho_{sr} ) \ln \frac{R}{R_{c} } $.
\item
  Layer 3:  $  ( \pm 1/3, \mp 1,  \pm 1 ) $ with energy $ E_{3}= E_{1} $ which is dictated by $ Z_{2} $ symmetry.
\end{itemize}

   The corresponding 2-body Kosterlize-Thouless (KT) transitions are
   $  k_{B} T^{2b}_{KT1}= k_{B} T^{2b}_{KT3} = \frac{ \pi}{2} ( \rho_{sl} +  \rho_{sr} ) $
   which is dictated by $ Z_{2} $ symmetry and $ k_{B} T^{2b}_{KT2} = 2 \pi \rho_{sr} $
   above which the QP and QH pairs are liberated into
   free ones. If setting $ \rho_{sl} = 3  \rho_{sr} $, then there is only one 2-body KT
   temperature $  k_{B} T^{2b}_{KT}= 2 \pi \rho_{sr} = \frac{ \pi \bar{\rho} }{ 9 m } $.
    As explained previously, microscopic calculations in the LLL \cite{un} are needed to 
   estimate the two spin stiffnesses $ \rho_{sl} $ and $ \rho_{sr} $.  
   In general, there should be two 2-body KT transition temperatures
   $  k_{B} T^{2b}_{KT1}= k_{B} T^{2b}_{KT3} \neq k_{B} T^{2b}_{KT2} $. 
   The spin wave excitations will turn into these Charge neutral two-body bound states
   at large wavevector  ( or short distance ). The two-body bound state behaves as a boson.
 
 (1) Two Charge $ {\pm 1} $ three-body bound states:

     The three-body bound states consists of three quasi-particles 
     $ ( \pm 1/3, \pm 1,  \pm 1 ), ( \pm 1/3, 0,  \mp 2 ), ( \pm 1/3, \mp 1,  \pm 1 ) $
     located at the three corners of a triangle ( Fig.1).
\vspace{0.5cm}

\epsfig{file=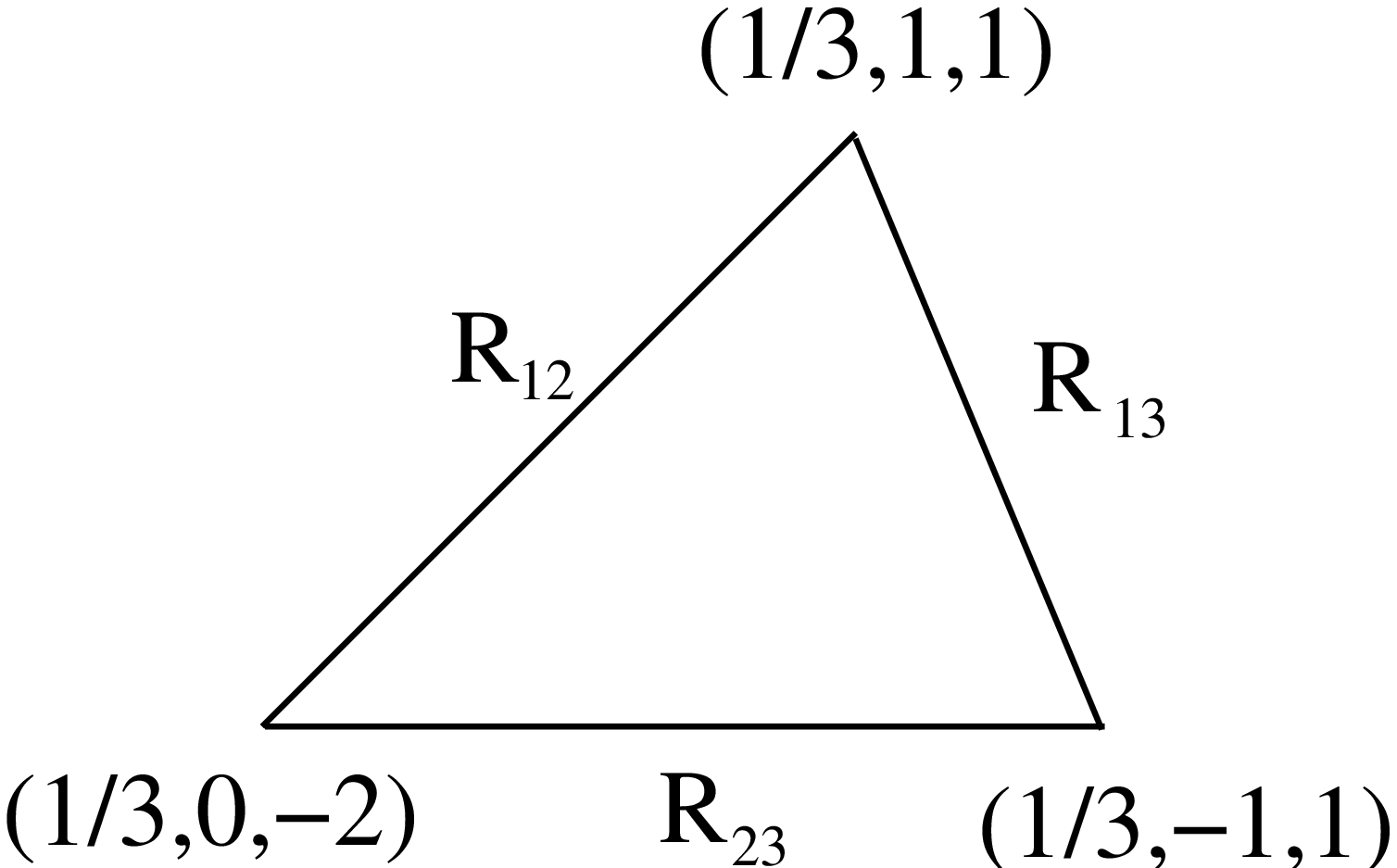,width=3.0in,height=1.5in,angle=0}

\vspace{0.25cm}

{\footnotesize {\bf Fig 1:} The lowest energy charged excitation is a three-body bound state
    of three $ \pm 1/3 $ charged quasi-particles sitting on the three corners of a triangle with
     $ R_{12} = R_{23} $ dictated by the $ Z_{2} $ symmetry. The three quantum numbers $ ( q, m_{l}, m_{r} ) $
     are enclosed in the parenthesis }

\vspace{0.25cm}

     Its energy  $ E_{3b}=  3E_{c \pm } + \frac{ e^{2} }{ 9 } (
     \frac{1} { R_{12} } + \frac{1} { R_{23} } + \frac{1} { R_{13} } ) +
     2 \pi  [ 2 \rho_{sr} \ln \frac{R_{12}}{R_{c}} + ( \rho_{sl} -  \rho_{sr} ) \ln \frac{R_{13}}{R_{c}}
     + 2 \rho_{sr} \ln \frac{R_{23}}{R_{c}} ]  $ when $ R \gg R_{c} \gg l $.
     Minimizing $ E_{3b} $ with respect to $ R_{12}, R_{13}, R_{23} $ leads to two optimal separations:
     $ R_{o12}= R_{o23} = \frac{ e^{2} }{ 36 \pi \rho_{sr} } $ which is dictated by the $ Z_{2} $ symmetry
     and $ R_{o13} = \frac{ e^{2} }{ 18 \pi ( \rho_{lr} - \rho_{sr} ) } $.
     If setting $ \rho_{sl} = 3  \rho_{sr} $, then the three quasi-particles are located
     on the corners of a isosceles with length $ R_{o} =  \frac{ e^{2} }{ 36 \pi \rho_{sr} } $,
     the corresponding 3-body Kosterlize-Thouless (KT) transition is the same as the two-body bound state
     $  k_{B} T^{3b}_{KT}= k_{B} T^{2b}_{KT} = 2 \pi \rho_{sr} = \frac{ \pi \bar{\rho} }{ 9 m } $.
     These charged
     excitations behave as fermions and are the main sources of dissipations in transport experiments.
     In general, $ T^{3b}_{KT} $ should be different from the two 2-body KT transition temperatures. 
    Again, microscopic calculations of the two spin stiffnesses $ \rho_{sl} $ and $ \rho_{sr} $ 
   in the LLL \cite{un} are needed to determine the three KT transition temperatures.

  Let's look at the interesting possibility of deconfined ( or free ) $ 1/2 $ charged excitations.
  Because any excitations with non-vanishing $ m_{l} $ or $ m_{r} $ will be confined,
  so any deconfined excitations
  must have $ m_{l} = m_{r}  =0  $ which implies $ m_{1} = m_{2} = m_{3} = m $ and $ m_{c} = 3m $.
  Eqn.\ref{charge} implies the charge $ q=  \frac{1}{3} m_{c}=m $ must be an integer. This proof rigorously
  rules out the possibility of the existence of deconfined fractional charges. We conclude
  that {\em any deconfined charge must have an integral charge}. $ m=1 $ corresponds to
  inserting one flux quantum through all the three layers which is conventional charge 1 excitation.
  Splitting the whole flux quantum into three fluxes penetrating the three layers at three  different
  positions will turn into the three-body bound state
  with the same charge shown in Fig.1. It is still not known if the energy of this conventional
  charge $ 1 $ excitation is lower than the three body bound state.

\section{ Broken symmetry}

 The broken symmetry and associated Goldstone modes can be easily seen from wavefunction
 approach.  In the first quantization, in the $ d \rightarrow 0 $ limit,
  the ground state trial wavefunction is the $ (111,111) $ state:
\begin{eqnarray}
  \Psi_{111,111} &  = &   \prod^{N_{1}}_{i=1} \prod^{N_{2}}_{j=1} (u_{i}-v_{j} )
    \prod^{N_{1}}_{i=1} \prod^{N_{3}}_{j=1} (u_{i}-w_{j} )
    \prod^{N_{2}}_{i=1} \prod^{N_{3}}_{j=1} (v_{i}-w_{j} ) \nonumber  \\ 
   & \times &  \prod^{N_{1}}_{ i< j } ( u_{i}-u_{j} )   \prod^{N_{2}}_{ i< j } ( v_{i}-v_{j} )
            \prod^{N_{3}}_{ i< j } ( w_{i}-w_{j} )
\label{ground}
\end{eqnarray}
    where $ u, v, w $ are the coordinates in layer 1, 2 and 3 respectively.

  In the second quantization, a trial wavefunction was written in \cite{hanna}:
\begin{eqnarray}
  | \Psi > & = &  \prod^{ M-1 }_{ m=0 } \frac{1}{ \sqrt{3} }
  ( e^{i \theta_{u} } C^{\dagger}_{m,1} + C^{\dagger}_{m,2} +
  e^{i \theta_{d} } C^{\dagger}_{m,3} ) |0>
                       \nonumber   \\
           & = &  \prod^{ M-1 }_{ m=0 } \frac{1}{ \sqrt{3} }
  ( 1 +  e^{i \theta_{u} } C^{\dagger}_{m,1} C_{m,2} +
  e^{i \theta_{d} } C^{\dagger}_{m,3} C_{m,2} )      \nonumber    \\ 
     & \times & \prod^{ M-1 }_{ m=0 } C^{\dagger}_{m,2} |0>  
\label{2nd}
\end{eqnarray}
  where $ \theta_{u}= \theta_{1} - \theta_{2}, \theta_{d} = \theta_{3} - \theta_{2} $
   and $ M=N $ is the angular momentum quantum number corresponding to the edge.
  We can interpret $ \theta_{u} $ mode in Eqn.\ref{2nd} as a "up" pairing between an electron
  in layer 1 and a hole in layer 2 which leads to a $ \theta_{u} $ molecule, similarly,
  $ \theta_d $ mode as a " down "  pairing between an electron in layer 3 and a hole
  in layer 2 which leads to a $ \theta_{d} $ molecule  ( Fig. 2).
\vspace{0.5cm}

\epsfig{file=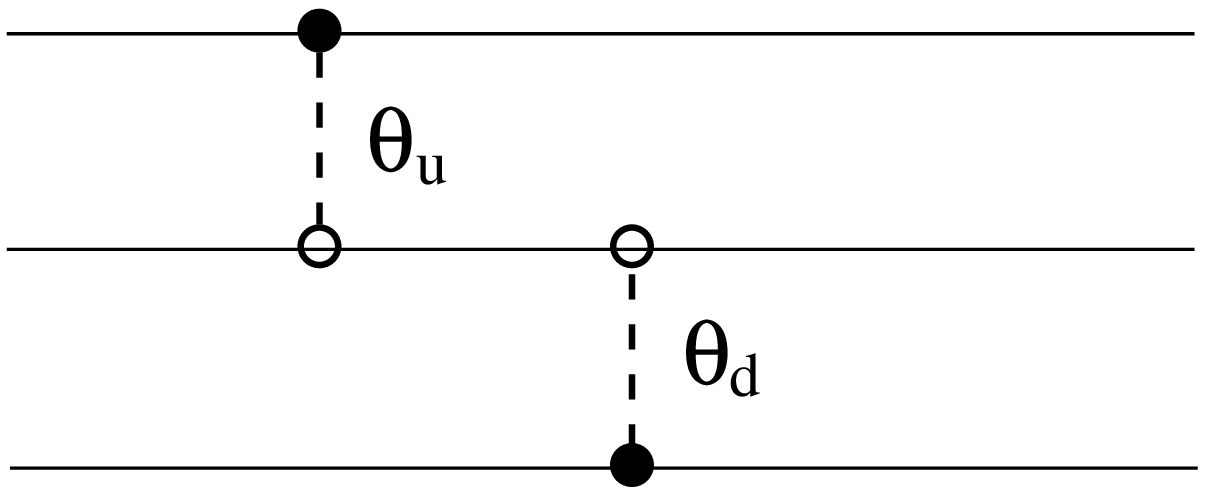,width=3.0in,height=1.5in,angle=0}

\vspace{0.25cm}

{\footnotesize {\bf Fig 2:} Two pairing modes of TLQH.
    Solid dots are electrons, open dots are holes. }

\vspace{0.25cm}

  Note that
  when $ \theta_{u}= \theta_{d} =0 $, $ | \Psi > $ is the ground state $ | G > $,
  it can be shown \cite{wave} that its projection on $ (N_{1}, N_{2}, N_{3} ) $ sector
  is $ (111,111) $ wavefunction in Eqn.\ref{ground}. 
  Note that Eqn.\ref{ground} is correct only in $ d \rightarrow 0 $ limit where the symmetry
  is enlarged to $ SU(3) $. In the $ SU(3) $ symmetric case, the pairing between any two layers of the three layers
  is equally likely. However, at any finite $ d $, Eqn.\ref{ground} and Eqn.\ref{2nd}
  may even not be qualitatively correct \cite{wave},
  because the $ SU(3) $ symmetry is reduced to $ U(1) \times U(1) \times U(1) \times Z_{2} $ symmetry
  ( where the three $ U(1) $ symmetry correspond to  $ c_{\alpha}(\vec{x}) \rightarrow
    e^{i \theta_{\alpha} } c_{\alpha}(\vec{x}), \alpha=1,2,3, c_{1} \leftrightarrow c_{3} $ )
  The pairings between any two layers out of the three layers are not equivalent any more.
  The two lowest energy pairing modes are shown in Fig. 2.

 $ SU(3) $ has $ 8 $ generators which can be written in terms of $ c_{\alpha} $:
\begin{eqnarray}
 S_{l} ( \vec{x}) & = & c^{\dagger}_{1}( \vec{x}) c_{1}( \vec{x}) - c^{\dagger}_{3}( \vec{x}) c_{3} ( \vec{x})
                 \nonumber  \\
 S_{r} ( \vec{x}) & = & c^{\dagger}_{1} ( \vec{x}) c_{1}( \vec{x})
     + c^{\dagger}_{3} ( \vec{x}) c_{3}( \vec{x}) - 2 c^{\dagger}_{2}( \vec{x}) c_{2}( \vec{x})
                    \nonumber  \\
 S_{12}( \vec{x}) & = & c^{\dagger}_{1}( \vec{x})c_{2}( \vec{x}),~~~~S_{21}( \vec{x})=
   S^{\dagger}_{12}( \vec{x})= c^{\dagger}_{2}( \vec{x})c_{1}( \vec{x})
                    \nonumber  \\
 S_{23}( \vec{x}) & = & c^{\dagger}_{2}( \vec{x})c_{3}( \vec{x}),~~~~S_{32}( \vec{x})= S^{\dagger}_{23}
 ( \vec{x})= c^{\dagger}_{3} ( \vec{x})c_{2}( \vec{x})
                    \nonumber  \\
 S_{13} ( \vec{x}) & = & c^{\dagger}_{1}( \vec{x})c_{3}( \vec{x}),~~~~S_{31}( \vec{x})=
  S^{\dagger}_{13}( \vec{x})= c^{\dagger}_{3}( \vec{x} )c_{1}( \vec{x} )
\label{cartan}
\end{eqnarray}
   where $ S_{l}, S_{r} $ are the two ( left and right )  generators in Cartan subalgebra (
   Note that the algebra is {\em not} $ SU(2) $ in spin $ j=1 $ representation which is also
   3 dimensional as used in \cite{fer} ). In the balanced case
  $ N_{1}=N_{2}=N_{3}=N/3 $, we have $ < S_{l} > = < S_{r} > = 0 $. The ground state in Eqn.\ref{ground}
  {\em breaks } two of the three $ U(1) $ symmetries, therefore there are two Goldstone modes in
  the corresponding two order parameters 
  $ < S_{12} >=  \frac{N}{3} e^{i \theta_{u} }, < S_{23} >=  \frac{N}{3} e^{ -i \theta_{d} } $.

  When $ \theta_{u} $ molecule and $ \theta_{d} $ molecule in Fig.2 move in the opposite direction,
  the currents in layer 2 cancel each other, it leads to the left-moving superfluid
  channel $ \theta_{l} $ ( Fig. 3a ).
  While when $ \theta_{u} $ molecule and $ \theta_{d} $ molecule move in the same direction,
  the currents in layer 2 add and flow in the opposite direction to the currents in layer 1 and
  layer 3, it leads to the right-moving superfluid
  channel $ \theta_{r} $ ( Fig. 3b ).
  In BLQH, there is only one superfluid channel which is the counter-flow channel \cite{em}. In TLQH, there
  are two superfluid channels which are left-moving and right-moving channels, the left-moving channel
  corresponds to the counter-flow channel \cite{note}.
  Obviously, it is easier to perform transport experiments in the counter-flow channel than
  in the right-moving channel.    
   
\vspace{0.5cm}

\epsfig{file=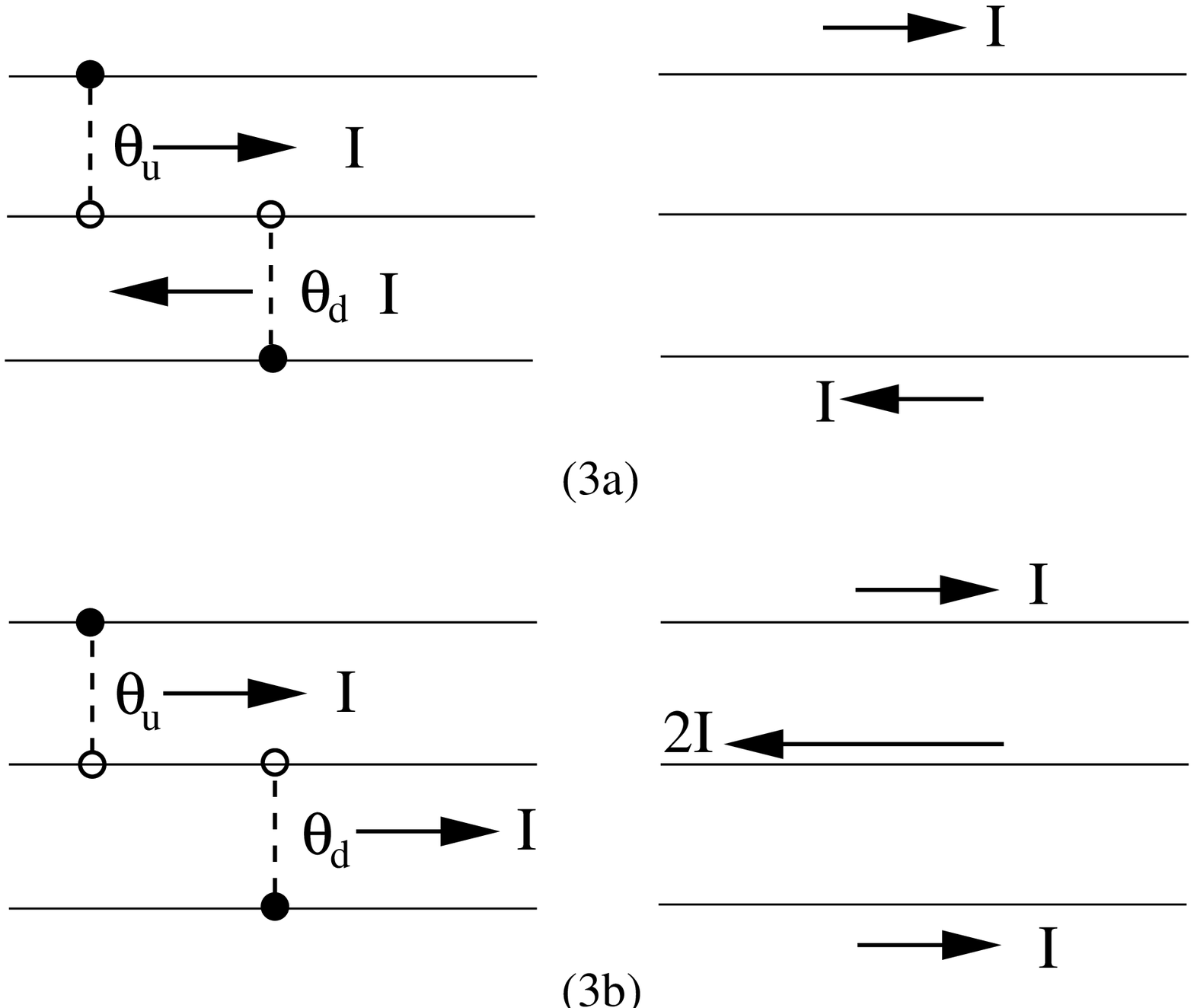,width=3.0in,height=2.0in,angle=0}

\vspace{0.25cm}

{\footnotesize {\bf Fig 3:} (3a) Left-moving superfluid channel which is the counter-flow channel
   (3b) Right-moving superfluid channel \cite{note} }

\vspace{0.25cm}
 
\section{ Correlated Interlayer tunnelings}

    So far, we assume that the tunnelings are tuned to be zero. In this section, we will
   outline several salient features in the tunnelings of TLQH.
    We assume the tunneling amplitude from layer 1 to layer 2 is the
    same as that of from layer 2 to layer 3.  The tunneling term is:
\begin{eqnarray}
    H_{t} & =  & t c^{\dagger}_{1} c_{2} + h.c. + t c^{\dagger}_{2} c_{3} + h.c.
    = t \phi^{\dagger}_{1} \phi_{2} + h.c. + t \phi^{\dagger}_{2} \phi_{3} + h.c.
              \nonumber  \\
   & =  & t \bar{\rho} ( \cos  \theta_{u}  + \cos \theta_{d}  )
   = 2 t \bar{\rho} \cos \frac{ \theta_{l} }{2} \cos \frac{ \theta_{r} }{2}
\end{eqnarray}
    where $ \theta_{u} = \theta_{1} - \theta_{2}, \theta_{d}= \theta_{3}-\theta_{2} $

    Integrating out the gapped charge sector $ \theta_{c} $ leads to the effective tunneling action:
\begin{eqnarray}
  {\cal L}  & =  &  \frac{1}{ 2 } [   \chi_{l}  ( \partial_{\tau} \theta_{l} )^{2} +  \rho_{sl}
      (  \nabla \theta_{l} )^{2} ] + \frac{ 1}{ 2 } [ \chi_{r} ( \partial_{\tau} \theta_{r} )^{2} +
      \rho_{sr} (  \nabla \theta_{r} )^{2} ]   \nonumber  \\
     & + & 2 t \bar{\rho} \cos \frac{ \theta_{l} }{2} \cos \frac{ \theta_{r} }{2}
\label{dep}
\end{eqnarray}
    where, as noted previously, the two spin susceptibilities $ \chi_{l}, \chi_{r} $
    and the two spin-stiffnesses $ \rho_{sl}, \rho_{sr} $ are  phenomilogical parameters
    to be fitted into the LLL+HF calculations or experimental data.

    Unfortunately, the above action is not very useful for any practical calculations, because
   the two angles $ \theta_{l}= \theta_{u} - \theta_{d}= \theta_{1} - \theta_{3},
  \theta_{r}= \theta_{u} + \theta_{d} = \theta_{1} - 2 \theta_{2} + \theta_{3} $ are
   not independent angles when considering the vortex excitations. 
   We can see this fact by noting that $  \theta_{l} + \theta_{r} = 2 \theta_{u} $
   or $  \theta_{l} - \theta_{r}= -2 \theta_{d} $
   has to be twice an angle which is a constraint between $  \theta_{l} $ and $ \theta_{u} $.
   While  $ \theta_{u} $ and $ \theta_{d} $ are two independent angles. So it is much more
   useful to write the above tunneling action in terms of the two
   independent angles $ \theta_{u}, \theta_{d} $:
\begin{eqnarray}
  {\cal L} & = & \frac{ 1 }{ 2 } ( \chi_{l} + \chi_{r} )
       ( \partial_{\tau} \theta_{u} )^{2} + \frac{ 1 }{ 2 } (  \rho_{sl} +
        \rho_{sr}  )(  \nabla \theta_{u} )^{2} + t \bar{\rho}\cos  \theta_{u}
                    \nonumber   \\
           & +  & \frac{ 1 }{ 2 } ( \chi_{l} + \chi_{r} )
       ( \partial_{\tau} \theta_{d} )^{2} + \frac{ 1 }{ 2 } (  \rho_{sl} + \rho_{sr}  )
       (  \nabla \theta_{d} )^{2} + t \bar{\rho} \cos \theta_{d}
         \nonumber   \\
    & + & ( \chi_{l} - \chi_{r} ) \partial_{\tau} \theta_{u}   \partial_{\tau} \theta_{d}
    + (  \rho_{sl} - \rho_{sr}  ) \nabla \theta_{u} \cdot \nabla \theta_{d}  
\label{tun3}
\end{eqnarray}
   where there is a $ Z_{2} $ symmetry under $ \theta_{u} \leftrightarrow  \theta_{d} $.
   The tunneling currents in three layers can be expressed in terms of the two independent angles:
   $ I_{1}= t\bar{\rho} \sin \theta_{u},
   I_{2}= - t \bar{\rho} ( \sin \theta_{u} + \sin \theta_{d} ), I_{3}=  t\bar{\rho} \sin \theta_{d} $.
   They satisfy the current conservation $ I_{1}+ I_{2} + I_{3} = 0 $.
      
   From Eqn.\ref{tun3}, we can see there is a coupling
    between $  \theta_{u} $ and $  \theta_{d} $.
     If there were not this coupling, we would have two independent BLQH tunneling actions. 
     It is this coupling which lead to the {\em correlated }
     tunnelings of TLQH.
     For example, the current out of layer 1 $ I_{1} = t\bar{\rho} \sin \theta_{u} $ only depends
     on $ \theta_{u} $ explicitly, while $ \theta_{u} $ couples to $ \theta_{d} $. 
     It was shown that the single gapless mode in BLQH leads to a sharp zero-bias peak
     in the interlayer tunneling of BLQH, the experimentally observed dissipations can only come from
     external sources such as disorder \cite{balents}. 
     If the coupling in TLQH even in the absence of external sources
     will lead to a broad zero-bias tunneling peak in TLQH will be investigated in \cite{un}.

    In the presence of in-plane magnetic field $ B_{||}=( B_{x}, B_{y} ) $, the gauge invariance dictates
    the tunneling term:
\begin{equation}
    H_{t} = t \bar{\rho} ( \cos ( \theta_{u} - Q \cdot x ) + \cos (\theta_{d} + Q \cdot x ) )
\end{equation}
    where $ Q_{\alpha} =  ( - \frac{ 2 \pi d B_{y} }{ \phi_{0} },
     \frac{ 2 \pi d B_{x} }{ \phi_{0} } )      $.

    In BLQH, it was found that the in-plane field will split the zero-bias peak, the splitting
   gives a direct measurement of the spin-wave velocity of the single Goldstone mode in
    BLQH \cite{balents}. However, in TLQH, 
   there are two coupled Goldstone modes with two different velocities $ v_{l}, v_{r} $. In \cite{un}, we will
   study how the zero-bias peak shifts and how to relate the shift to the two spin wave
   velocities or their ratio in the presence of in-plane magnetic field.

     We can also see that if there were not the coupling between $  \theta_{u} $ and $  \theta_{d} $,
   we would have two independent  Pokrovsky-Talapov (PT) models. In BLQH which is described by a single PT model,
   it was found that when the applied in-plane magnetic field
   is larger than a critical field $ B > B^{*}_{||} $,
   there is a phase transition from a commensurate state to an incommensurate state (C-IC)
   with broken translational symmetry. When $ B > B^{*}_{||} $, there is a finite temperature KT transition
   which restores the translation symmetry by means of dislocations
   in the domain wall structure in the incommensurate phase \cite{yang,rev}.
   In TLQH which is described by the above two coupled PT model \cite{ising}, several kinds of
   C-IC transitions and associated several kinds of KT transitions are expected and will
   be studied in a separate publication \cite{un}. 

\section{ Conclusions:} 

 In this paper, we study the interlayer coherent incompressible phase
 in spin polarized Trilayer Quantum Hall systems at total filling factor
 $ \nu_{T}=1 $ from three approaches: MCF, CB and wavefunction approach.
 Just like in TLQH, CB approach is superior
 than MCF approach. 
 The Hall and Hall drag resistivities are found to be quantized at $ h/e^{2} $.
 Two neutral gapless modes which are left and right moving modes
 with linear dispersion relations $ v_{l/r} = \omega_{l/r} k $ are identified and
 the ratio of the two velocities is close to $ \sqrt{3} $.
 The novel excitation spectra are classified into two classes:
 (1) Charge neutral bosonic  two-body bound states.
 (3) Charge $ \pm 1 $ fermionic three-body bound states.
  In general, there are two 2-body KT transition temperatures and one 3-body KT transition
  temperature. Microscopic calculations 
   in the LLL \cite{un} are needed to roughly estimate the three KT transition temperatures.
 The Charge $ \pm 1 $ three-body bound states shown in Fig. 1 may be the main
 dissipation source of the interlayer tunneling.
 The broken symmetry in terms of $ SU(3) $ algebra is studied.
 The effective action of  the two coupled 
 Goldstone modes are derived. The  excitonic structure in TLQH
 was shown in Fig. 2.
  In BLQH, there is only one superfluid channel which is the counter-flow channel \cite{em}. In TLQH, there
  are two superfluid channels which are left-moving and right-moving channels ( Fig.3 ), the left-moving channel
  corresponds to the counter-flow channel ( Fig. 3a ).
  Obviously, it is easier to perform transport experiments in the counter-flow channel than
  in the right-moving channel. In the presence of interlayer tunnelings, the coupling between the two Goldstone modes
  may lead to the broadening of the zero-bias tunneling peak,
  in contrast to the zero-bias peak in BLQH. The precise nature of the broadening
  and several other possible new phenomena unique to the TLQH in the presence of in-plane magnetic field
  will be studied in a forthcoming publication.

   We expect that the effective action of the two coupled Goldstone modes
  and the qualitative physical pictures 
  of TLQH achieved from CB approach in this paper are correct.
  However, the parameters in the effective action need to be estimated from microscopic calculations
  in the LLL \cite{un}. It is interesting to see if the ratio $ v_{l}/v_{r} \sim \sqrt{3} $ still holds in
  the microscopic calculations. As stressed in \cite{moon,wave} for BLQH, the trial wavefunction
  Eqn.\ref{ground} and Eqn.\ref{2nd} may not even be qualitatively correct in TLQH, so it is still
  difficult to estimate the ratio with controlled accuracy even in the LLL+HF approach. Eventually,
  the ratio need to be measured by experiments.

    I thank J. K. Jain for helpful discussions.

\end{multicols}

\begin{references}
\bibitem{note} The "left" and "right " channels are used just to distinguish the two gapless channels.
               They have nothing to do with the moving directions in the real space.
\bibitem{ising} This relation resembles that between Ising model and Ashkin-Teller model
    which  is a model of two coupled Ising models.
\end{references}
\end{document}